# Artificial Ant Species on Solving Optimization Problems

Camelia-M. Pintea[a]

Technical University Cluj Napoca, North University Center Baia Mare, Romania

**Abstract.**

During the last years several ant-based techniques were involved to solve hard and complex optimization problems. The current paper is a short study about the influence of artificial ant species in solving optimization problems. There are studied the artificial *Pharaoh Ants, Lasius Niger* and also artificial ants with no special specificity used commonly in *Ant Colony Optimization*.

## 1 Introduction

*Ant Colony Optimization (ACO)* is nowadays one of the most efficient bio-inspired techniques for solving complex combinatorial optimization problems. Introduced by Dorigo, the ant-based metaheuristic proves over time the ability on solving many difficult theoretical and real-life problems. At first was used to solve routing problems as the *Traveling Salesman Problem (TSP)* [1] and with time more and more *Combinatorial Optimization Problems (COP)* were solved based on this bio-inspired technique.

Different versions of *Traveling Salesman Problem* were solved using ant system; for example the *Railway Traveling Salesman Problem* [6], *Generalized Traveling Salesman Problem* [17,25] and *Fuzzy TSP* with ant algorithms [2,3].

The component-based models for solving *TSP* with ant-based models are described in [22]. Some examples of other complex problems solved by the author with hybrid ant models are the *Linear Ordering Problem* [5,16], the *Gate Assignment Problem* [23], the *Matrix Bandwidth Problem* [26] and the *Generalized Vehicle Routing Problem* [27].

Dynamic real-life problems are more difficult to solve than the static ones. Several author's approaches based on artificial ants used to solve some dynamic problems are the parallel *ACO* with a ring neighbourhood for *Dynamic TSP* [10], an ant colony algorithm for solving the *Dynamic Generalized Vehicle Routing Problem* [12], an ant-based technique for the *Dynamic Generalized Traveling Salesman Problem* [21] and the bio-inspired approach for *Dynamic Railway Problem* [28].

The current paper is a study of several types of ants used as models for their artificial homonyms. It is analyzed the influence of the specific characteristics of artificial ants on solving combinatorial optimization problems. Section two describes how are used in general the artificial ants on solving optimization problems, followed by the sections where are involved the artificial *Lasius Niger* ants and further more the *Pharaoh ants* on solving combinatorial optimization problems.

## 2 The Artificial Ants on Solving Optimization Problems

The *Ant Colony Optimization* is based in general on the ant's ability on finding the shorter paths between their nest and the food location involving social learning [11,14].

The artificial ants are using the indirect communication based on pheromone quantity laid on their trails. As in real life, the ants use the trails with a strong pheromone showing that the most promising tour is the one with higher amounts of pheromone. The artificial ants are endowed with an initial quantity of virtual pheromone and the algorithm is challenging with updating ants' pheromones. In the first *ACO* model, *Ant System (AS),* was used a global updating pheromone rule to improve the already best found tour solution.

Email address: cmpintea @ yahoo.com (Camelia-M. Pintea)

Furthermore, in *Ant Colony System (ACS)* and in many other variants of *ACS* was introduced also a local updating pheromone rule in order to favor exploration and as a consequence to find a better local solution. An inner update rule described in [17] is used to improve the ant-based search in the local neighborhood. In order to enforce the construction of a valid solution used in *ACS* new algorithms were elaborated using the inner rule and tested on several complex problems [17].

A brief *ACS* description for solving *TSP* follows. At first the ants are positioned on graph - the representation of the *TSP* - vertices chosen according to some initialization rule (e.g., randomly). Each ant builds a feasible solution by repeatedly applying a state transition rule. Meanwhile the ant modifies the amount of pheromone on the traversed edges by applying the local updating rule. When all ants completed the tours, the amount of pheromone on edges is modified by the elitist ants - the ants that found the best tour - applying the global updating rule [1].

In [8] a hybrid *ACS* for solving the *Matrix Bandwidth Minimization Problem* (*MBMP*), including local procedures, based on the *Cuthill-McKee* method [9], is described. *Cuthill-McKee* method is about swapping the vertices with the highest degree with randomly selected ones with minimum degrees. *MBMP* is about seeking a permutation of the rows and columns of a symmetric matrix such that the non-zero elements are as close as possible to the main diagonal.

A lot of ants find "not-so-good" solutions, but the repeatedly swapping technique "encourages" few elitist ants to find the best solution. The hybrid ant swapping approach performed better than *ACS* on large instances when using the same parameters.

## 3 The Artificial Pharaoh Ants on Solving Optimization Problems

*Pharaoh ants* are originally from Africa and are also called *Monomorium pharaonis*. The worker ants are only two mm long and as the labs study shows they are not guided by their memory or landmarks when foraging.

As a specific characteristic, the *Pharaoh ant* colonies are not well defined, they are called "unicolonial"; it is easy to add or remove ants on create or remove colonies to whatever size [18]. In general, the trail persistence depends on pheromone longevity, the number of ants, the intensity of trail lying by each ant and also depends on the environmental conditions. The pheromone duration of *Monomorium pharaonis* is close to 30 minutes.

On laboratory study in the branch choice experiment, trail pheromones of *Pharaoh ant* decay rapidly. The results show that 37% of the ants make a *U-turn* and also walk with their sting extended. The results indicate a behavioural specialization: several ants are walking backwards and forwards laying pheromone, for guiding the other ants [13,24]. It is required a continual flow of workers to replace any pheromone that decays; therefore are necessary a large number of ants to extend the exploitation.

In Pintea et al. 2007 [19] is introduced the *Pharaoh Ant System (PAS)*. *Pharaoh Ant System* is based on *Ant Colony System*. As was already specified, it is necessary to use a lot of ants; therefore three ant colonies are involved. Two ant colonies are used for explore the solutions space but also one colony have the role on spreading the negative pheromone on bad trails. The third colony is exploiting the solutions domain provided by the first two colonies. In this phase, the third colony use the 'no-entry' signal. As the real pharaoh ants, the artificial ones are making u-turns to establish the best tour on the paths.

In [15] are described the component models with *Pharaoh System* for the labyrinth problem. Based on *AntNet* [7] and *Pharaoh Ant System* [19] was introduced the *Distributed Pharaoh System (DPS)* [20]. *DPS* is using three ant colonies with the same functions and characteristics as in *PAS*. The forward ants, the explorers, construct a path from the source to the destination traversing the network. The backward ants are the exploiters and are retracing the path updating pheromone tables and routing information; also zigzagging or making *U-turns* when founds bad trails. This is how it is achieved the *DPS* objective: finding a low cost overlay network topology.

## 4 The Artificial Lasius Niger Ants on Solving Optimization Problems

The common garden ant, *Lasius niger*, also uses trail pheromones. In this species the trail seems to last longer than in *Pharaoh's ants*. Experienced foragers seem to rely mainly on route memory not pheromone trails in relocating a feeding site. Possibly, the role of trail pheromone is more to help naive ants in locating food [13,24].

*Lasius niger* ants behaviour [11] inspired the *Step Back Sensitive Ant Model (SB-SAM)* from [5]. The step-back ant-model is an extension of *Sensitive Ant Model (SAM)* [4] including a virtual state for ants with a certain sensitivity level. The virtual decision rule is associated with *Lasius niger u-turns*, studied by biologists. In biological studies is showing that *Lasius niger* ants make *u-turns* and exploits the geometry of the trail network bifurcation [11].

The virtual state transition rule from *SB-SAM* take a step back when select the previous node [5]. The *u-turns* give a better path when compared to bidirectional trail laying. The step-back induces diversity in the search; the pheromone trail decreases on the edge connecting the current node with the previous one. To find best solutions, it is important keep a balance between exploitation and exploration [5] and also a good distribution of pheromone sensitivity levels in the ant population.

The *Pharaoh ants* and *Lasius niger* have some particular characteristics and influence in their specific way the ant-based algorithms for solving complex optimization problems.

## 5 Conclusions

The indirect communications between agents in bio-inspired algorithms is well represented in ant-based models. *Ant Colony Optimization* is using artificial pheromone trails when artificial ants build solutions within the problem's graph representation. In ant systems it is important to know how long the artificial pheromone is deposited on the trail, based on the ants' specific characteristics.

Based on several biological studies *Lasius niger* 's pheromone trail last longer than in *Pharaoh's ants.* Several problems solved with artificial ants including *Pharaoh ants*, *Lasius niger* and common ants are described, including the way the characteristics of ants influence the problems results.

**Bibliography**

1. Dorigo, M., Stützle, T. - *Ant Colony Optimization*, MIT Press (2004)
2. Crisan, G-C, Nechita, E.: Solving Fuzzy TSP with Ant Algorithms, Int J Comput Commun 3(S): 228–231 (2006)
3. Crisan, G-C.: Ant Algorithms in Artificial Intelligence, PhD Thesis, "Al. I. Cuza" University of Iasi (2007)
4. Chira C, Pintea C-M, Dumitrescu D.: Heterogeneous Sensitive Ant Model for Combinatorial Optimization. GECCO 2008. ACM 163–164 (2008)
5. Chira, C., Pintea, C.-M., Crisan, G-C., Dumitrescu, D.: Solving the Linear Ordering Problem using Ant Models, GECCO 2009, ACM, 1803-1804 (2009)
6. Pop, C.P., Pintea, C-M., Sitar, C.P.: An Ant-based Heuristic for the Railway Traveling Salesman Problem, LNCS 4448, 702-711 (2007)
7. Caro, G.D., Dorigo, M.: Antnet: Distributed stigmergetic control for communications networks. J. Artificial Intelligence Research, 9: 317-365 (1998)
8. Crisan, C., Pintea, C-M.: A hybrid technique for a matrix bandwidth problem, Sci. Stud. Res., Math.Inform. 21(1): 113-120 (2011)
9. Cuthill, E., McKee J., Reducing the bandwidth of sparse symmetric matrices, In Proceedings of the 24th National Conference ACM. 157-172 (1969)
10. Pintea, C-M., Crisan, G-C., Manea, M.: Parallel ACO with a Ring Neighbourhood for Dynamic TSP, J. Information Technology Research, 5(4):1-13 (2012)
11. Beckers R, Deneuborg J L, Goss S Trails and U-turns in the Selection of a Path of the Ant Lasius Niger. J Theoretical Biology. 159:397–415 (1992)

12. Pop, P.C., Pintea, C-M., Dumitrescu, D.: An Ant colony algorithm for solving the Dynamic Generalized Vehicle Routing Problem, Civil Engineering 1(11):373-382 (2009)
13. Grüter, C., Czaczkes, T. J., Ratnieks, F. L.: Decision making in ant foragers (Lasius niger) facing conflicting private and social information. Behavioral Ecology & Sociobiology, 65(2): 141-148. (2011)
14. Grüter, C., Leadbeater, E., Ratnieks, F. L. W.: Social learning: the importance of copying others. Current Biology 20: R683-R685. (2010)
15. Pintea, C-M., Vescan, A.: The Labyrinth Problem: Component Model with Pharaoh System, Conference ICFS 2007, Ed.Oradea University, 82-86 (2007)
16. Pintea, C.-M., Chira, C., Dumitrescu, D.: New results of ant algorithms for the Linear Ordering Problem, An. Univ. Vest Timis., Mate-Info.48(3): 139-150 (2010)
17. Pintea, C-M.: Combinatorial optimization with bio-inspired computing. PhD Thesis, Babes-Bolyai University (2008)
18. Jeanson, R.L., Ratnieks, F.L.W., Deneubourg, J-L.: Pheromone trail decay rates on different substrates in the Pharaoh's ant, Monomorium pharaonis. Physiological Entomology 28:192-198 (2003)
19. Pintea, C-M., Dumitrescu, D.: Introducing Pharaoh Ant System, MENDEL 2007 Brno University, 54-59 (2007)
20. Pintea, C-M., Dumitrescu, D.: Distributed Pharaoh System for Network Routing, Automat. Comput. Appl. Math.16(1-2):27-34 (2007)
21. Pintea, C-M. Pop, P.C. Dumitrescu, D.: An Ant-based Technique for the Dynamic Generalized Traveling Salesman Problem, 7-th Int. Conf. on Systems Theory and Scientific Computation, 257-261 (2007)
22. Vescan, A., Pintea, C-M.: Component-based Ant Systems, Modern Paradigms in Computer Science and Applied Mathematics. AVM, München (2011)
23. Pintea, C-M., Pop, P.C., Chira, C., Dumitrescu, D.: A Hybrid Ant-based System for Gate Assignment Problem, LNCS 5271, 273-280 (2008)
24. Evison, S. E. F., Petchey, O. L., Beckerman, A. P., Ratnieks, F. L. W.: Combined use of pheromone trails and visual landmarks by the common garden ant Lasius niger. Behavioral Ecology & Sociobiology 62: 261-267 (2008)
25. Reihaneh, M., Karapetyan, D.: An efficient hybrid ant colony system for the generalized traveling salesman problem. Algorithmic Operational Research, 7, 21-28 (2012)
26. Pintea, C-M., Crisan, G-C, Chira, C.: A Hybrid ACO Approach to the Matrix Bandwidth Minimization Problem, LNCS 6076, 405-412 (2010)
27. Pintea, C-M., Chira, C., Dumitrescu, D., Pop, P.C.: Sensitive Ants in Solving the Generalized Vehicle Routing Problem, Int J Comput Commun, 6,4,731-738 (2011)
28. Pop, C.P., Pintea, C-M., Sitar, C.P., Dumitrescu, D.: A Bio-inspired Approach for a Dynamic Railway Problem, SYNASC 2007, 449-452 (2007)